\documentclass[fleqn,12pt,twoside]{article}
\usepackage[headings]{espcrc1}

\readRCS
$Id: espcrc1.tex,v 1.2 2004/02/24 11:22:11 spepping Exp $
\usepackage{graphicx}
\usepackage[figuresright]{rotating}

\newcommand{\AmS}{{\protect\the\textfont2
  A\kern-.1667em\lower.5ex\hbox{M}\kern-.125emS}}

\title{Ultra-peripheral collisions of heavy ions at RHIC and the LHC}

\author{Joakim Nystrand\address{Department of Physics and Technology, 
        University of Bergen, 
        All{\'e}gaten 55, N-5007 Bergen, Norway}
}
       
\runtitle{Ultra-peripheral collisions of heavy ions at RHIC and the LHC}
\runauthor{J. Nystrand}

\begin{document}

\maketitle

This paper deals with so-called Ultra-Peripheral Collisions (UPCs) of heavy 
ions\cite{Bertulani:2005ru,Baur:2001jj}.  
These can be defined as collisions in which no hadronic 
interactions occur because of the large spatial separation between the 
projectile and target. The interactions are instead mediated by the 
electromagnetic field. 
Two types of ultra-peripheral collisions can be distinguished: purely 
electro-magnetic interactions (two-photon interactions) and photonuclear 
interactions, in which a photon from 
the projectile interacts with the hadronic component of the target.

\section{Electromagnetic interactions at hadron colliders}

Photon-induced interactions have traditionally been studied with electron 
beams. 
In collisions between hadrons, the strong interaction will always dominate 
at all reasonable energies, and the cross section for electromagnetic particle 
production will only be a small fraction of the cross section for the corresponding 
strong processes. 

Despite this, there are several reasons to study electromagnetic interactions at 
hadron colliders. First and foremost, with the 
construction of the Large Hadron Collider (LHC) at CERN, the range of accessible photon 
energies will be strongly increased and the equivalent luminosities will be higher 
than at existing electron colliders. 
Furthermore, with nuclear beams, effects of very strong fields can be 
studied (the effective coupling is enhanced by the nuclear charge Z to 
$Z \sqrt{\alpha}$ rather than just $\sqrt{\alpha}$)\cite{Baur:2003ar}. 
The fact that both beam particles 
can act as photon emitter and target in a symmetric collision ($pp$ or 
AA) leads to some interesting interference effects, not present 
in e-p or e-A collisions\cite{Klein:1999gv}. 

\section{The equivalent photon luminosity}

As was realized by Fermi already in 1924, before the first formulation 
of a quantum field theory, and further elaborated by Weizs\"{a}cker and Williams some 
10 years later, the electromagnetic field 
of a relativistic particle corresponds to an equivalent flux of photons. 
The spectrum of photons with energy $k = x E_{beam}$ and virtuality 
$q^2 = - Q^2$ associated with a point particle of charge $eZ$ is given by\cite{Budnev:1974de}
\begin{equation}
\label{dndxdq2}
x \frac{dn_{\gamma}}{dxdQ^2} = \frac{\alpha Z^2}{\pi} \, (1 - x + 1/2 x^2) 
\frac{Q^2 - Q_{min}^2}{Q^4} \; .
\end{equation}
The photon virtuality is equal to the 4-momentum transfer from the projectile, 
$q^2 = (p_i - p_i')^2$, in a scattering reaction, and $Q^2_{min} = (x m)^2/(1-x)$, 
where $m$ is the mass of the projectile. 
The total number of equivalent photons of a given energy is obtained by integrating 
Eq.~\ref{dndxdq2} from $Q^2_{min}$ to some $Q^2_{max}$. This gives 
\begin{equation}
\label{dndx}
x \frac{dn_{\gamma}}{dx} = \frac{\alpha Z^2}{\pi} \, (1 - x + 1/2 x^2) 
\left[ \ln( \frac{Q^2_{max}}{Q_{min}^2} ) + \frac{Q_{min}^2}{Q^2_{max}} -1 \right] \; .
\end{equation}
The theoretical maximum $Q^2_{max}$, corresponding to a scattering angel $\theta = 180^o$, 
is $Q^2_{max} = 4 E_{beam}^2 (1-x)$. In practice, the maximum momentum transfer is often 
limited by experimental constraints on the maximum scattering angle. 

For an extended target, such as a proton or a nucleus, the $Q^2$ dependence in Eq.~\ref{dndxdq2} 
is modified by a form factor, $| F(Q^2) |^2$. This provides a natural cut-off for 
high values of $Q^2$. The photon spectrum associated with a proton, assuming a 
dipole form factor, $F_E (Q^2) = 1/(1 + Q^2/0.71 GeV^2)^2$ is\cite{Nystrand:2004vn}  
\begin{equation}
\label{JN}
x \frac{dn_{\gamma}}{dx} = \frac{\alpha}{\pi} \, (1 - x + 1/2 x^2)
\left[ \frac{A+3}{A-1} \ln(A) - \frac{17}{6} - \frac{4}{3A} + \frac{1}{6 A^2} \right]  \; ,
\end{equation}
where $A = 1 + (0.71 GeV^2)/Q_{min}^2$. The photon spectrum 
of high energy protons is also discussed in\cite{Khoze:2002dc}.   

The photon spectrum associated with a single nucleus can be well described by 
introducing a nuclear form factor. However, to realistically describe the photon 
spectrum in an interaction between {\it two} nuclei, a better method is to calculate 
the spectrum as function of impact parameter in a semi-classical approach. In this way, 
interactions where the nuclei interact strongly can be excluded (determined roughly 
by $b_{min} > 2R$)\cite{Baur:2001jj}.  

\begin{table}[htb]
\caption{Summary of accelerator parameters, from \cite{Yao:2006px}. $E_{beam}$ is the energy 
of one of the beam particles in the rest frame of the other (electron energy in the rest frame 
of the proton for HERA). 
The numbers for nucleus-nucleus collisions are per nucleon or per nucleon-nucleon collision.
The luminosities are peak luminosities.}
\label{accelerators}
\newcommand{\m}{\hphantom{$-$}}
\newcommand{\cc}[1]{\multicolumn{1}{c}{#1}}
\renewcommand{\tabcolsep}{1.15pc} 
\renewcommand{\arraystretch}{1.2} 
\begin{tabular}{@{}lllll}
\hline
Accelerator    & System   & Luminosity [cm$^{-2}$ s$^{-1}$] & $\sqrt{s}$ & $E_{beam}$ [GeV] \\ \hline
RHIC           & Au+Au    & $1.5 \cdot 10^{27}$   & 200 GeV    & $2.1 \cdot 10^{4}$ \\
HERA--II       & e+p      & $7.5 \cdot 10^{31}$   & 318 GeV    & $5.4 \cdot 10^{4}$ \\
LHC            & p+p      & $1.0 \cdot 10^{34}$   & 14  TeV    & $1.0 \cdot 10^{8}$ \\
LHC            & Pb+Pb    & $1.0 \cdot 10^{27}$   & 5.5 TeV    & $1.6 \cdot 10^{7}$ \\
\hline
\end{tabular}\\[2pt]
\end{table}

\begin{figure}
\begin{center}
\includegraphics[width=10cm]{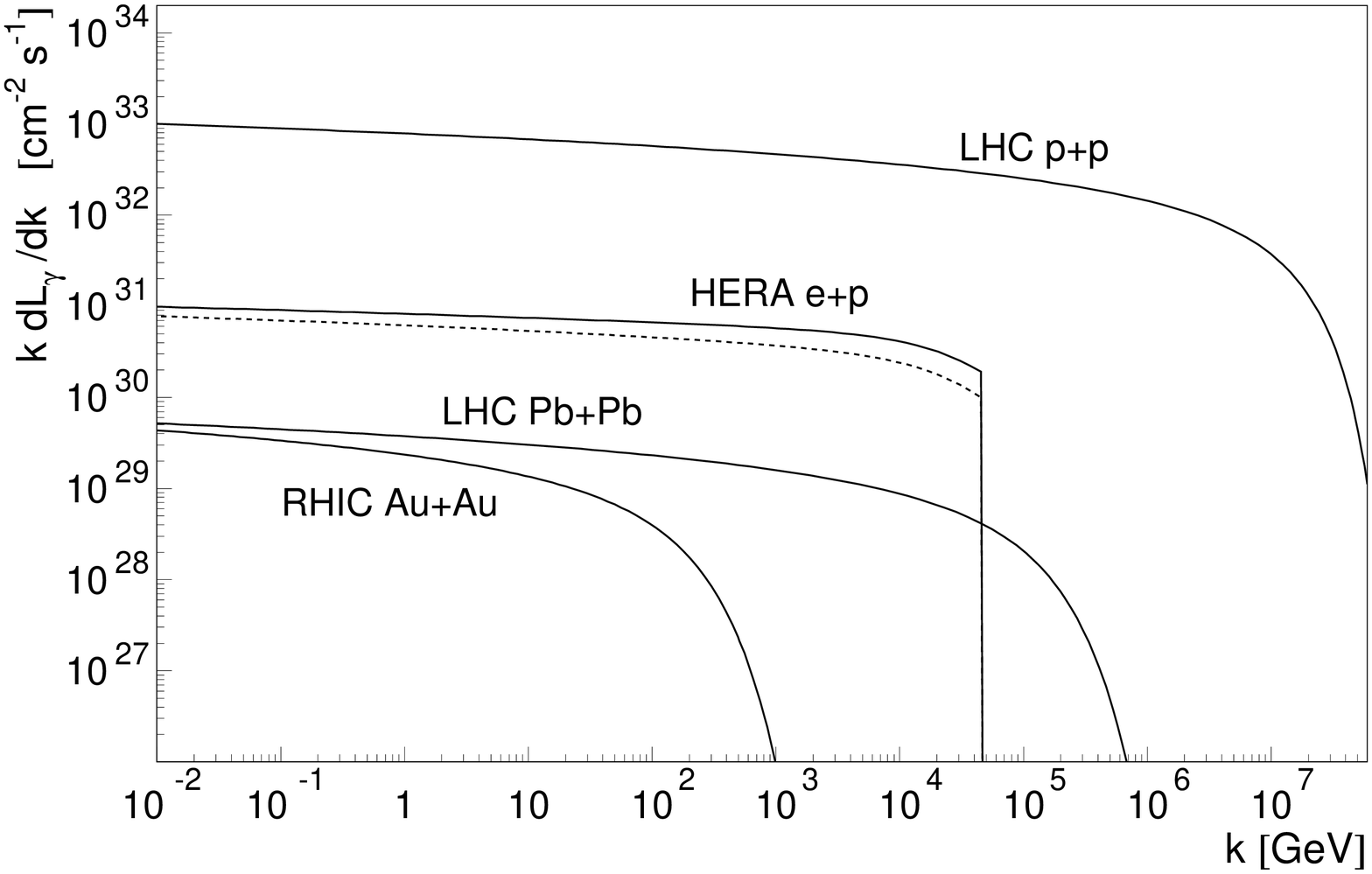}
\caption{A comparison of the equivalent photon luminosity at HERA, RHIC and the LHC. 
Here, k is the photon energy in the target rest frame. The beam energies and luminosities 
are taken from Table~1. The solid curve for HERA corresponds to the theoretical maximum 
of $Q^2_{max}$, whereas the dashed curve corresponds to $Q^2_{max} = m_{\rho}^2$, appropriate 
for hadron production\cite{Budnev:1974de}.}
\label{photon_spectrum}
\end{center}
\end{figure}

A comparison of the equivalent photon luminosities at hadron colliders and at the 
electron-proton collider HERA at DESY is shown in Fig.~\ref{photon_spectrum}. 
The comparison is made for $pp$ and Pb+Pb interactions at the LHC and for Au+Au 
interactions at the Relativistic Heavy-Ion Collider (RHIC) at Brookhaven National Laboratory. 
The photon luminosity 
is defined as the photon spectrum, $dn_{\gamma}/dk$, multiplied by the beam luminosity. 
The machine luminosities and beam energies are listed in Table~\ref{accelerators}.  
The photon energies at the LHC, both in $pp$ and Pb+Pb interactions, 
extend far beyond the maximum at HERA. 
A similar comparison between 
the photon flux in heavy-ion  interactions at RHIC and the LHC with the expected flux 
at eRHIC was made in\cite{White:2001tc}, but a very low cut-off in $Q^2_{max}$ for 
electrons ($Q^2_{max} < 4 m_e^2$) was used there.

\section{Particle production in  ultra-peripheral collisions} 

For nuclear beams, the form factor cuts off the equivalent photon energy spectrum 
above $\sim \gamma / R$, and the photon flux is significant only well below this energy. 
This cut-off in photon energy corresponds to maxima in the photon-nucleon center 
of mass energies of $W_{\gamma p} \approx 35$~GeV and $W_{\gamma p} \approx 950$~GeV 
in Au+Au/Pb+Pb collisions at RHIC and the LHC, respectively. 
These energies are sufficient for producing heavy particles. 

As an illustration of the  
different production mechanisms in central and ultra-peripheral collisions, consider the production 
of heavy quark pairs, $c \overline{c}$ and  $b \overline{b}$. 
The dominant partonic process in hadronic interactions is gluon-gluon fusion   
$g+g \rightarrow c \overline{c}$ ($q \overline{q} \rightarrow c \overline{c}$ also 
contributes at lowest order in $\alpha_s$, but gives only a small 
contribution for energies far above the threshold $2m_Q$)\cite{Vogt:2001nh}. 
The total cross section for 
$pp$ collisions can be calculated from the parton distribution functions. If nuclear 
shadowing is neglected, and binary scaling assumed, the total cross section 
(integrated over all centralities) 
in a nucleus-nucleus collision at the same energy will be 
$A^2 \sigma(pp \rightarrow Q\overline{Q} + X)$. $A$ is the nuclear mass number. 

The corresponding interaction in photonuclear collisions is 
$\gamma + g \rightarrow Q \overline{Q}$\cite{Klein:2002wm}. The cross section for 
this process can 
be similary calculated from the equivalent photon spectrum and the nuclear gluon 
distribution function. Table~\ref{quarkpairs} compares the cross sections for 
the two production mechanisms in Pb+Pb collisions at the LHC. 

\begin{table}[htb]
\caption{Comparison of the cross sections for charm- and bottom-quark pairs through 
hadroproduction (dominated by $g+g \rightarrow Q\overline{Q}$), 
photoproduction ($\gamma+g \rightarrow Q\overline{Q}$), and 
two-photon production ($\gamma+\gamma \rightarrow Q\overline{Q}$) in Pb+Pb interactions 
at the LHC. The photoproduction and two-photon calculations are from\cite{Klein:2002wm}. 
The hadroproduction cross sections are calculated as $A^2 \sigma(pp \rightarrow Q\overline{Q})$ 
with $\sigma(pp \rightarrow Q\overline{Q})$ from\cite{Vogt:2001nh} (NLO calculations).}
\label{quarkpairs}
\newcommand{\m}{\hphantom{$-$}}
\newcommand{\cc}[1]{\multicolumn{1}{c}{#1}}
\renewcommand{\tabcolsep}{0.27pc} 
\renewcommand{\arraystretch}{1.2} 
\begin{tabular}{@{}llll}
\hline
quarks       & hadroproduction           & photoproduction   & two-photon production \\ 
             & {\small $\sigma(Pb+Pb \rightarrow Q\overline{Q}+X)$} & {\small $\sigma(Pb+Pb \rightarrow Pb+Q\overline{Q}+X)$} & {\small $\sigma(Pb+Pb \rightarrow Pb+Pb+Q\overline{Q})$} \\ \hline 
$c \overline{c}$ & 252 b$^1$                 & 1.2 b               & 1.1 mb \\
$b \overline{b}$ & 8.1 b$^1$                 & 4.9 mb              & 0.9 $\mu$b \\
\hline
\end{tabular}\\[2pt]
$^1$ The hadroproduction cross sections are larger than the geometrical cross section since 
multiple $Q \overline{Q}$--pairs can be produced in central collisions. 
\end{table}

The cross sections for photoproduction are about 
2-3 orders of magnitude smaller than those for hadroproduction. This is a consequence 
of the different coupling strengths of the strong and electromagnetic interactions, and the 
cut-off in the photon spectrum imposed by the coherence requirement. On an absolute scale, 
however, the cross section for photoproduction of $c \overline{c}$-pairs ($\approx$ 1~b) 
is by no means small compared with e.g. the geometrical cross section of about 6 b. 

A third mechanism by which quark pairs can be produced is through a two-photon interaction. 
From the cut-off in the photon energy, 
the maximum invariant mass of the produced state is then limited to 
$\sim 2 \gamma /R$. This is 6 and 160~GeV for heavy nuclei at RHIC and the LHC, respectively. 
The energy is thus sufficient for production of both $c \overline{c}$- and 
$b \overline{b}$-pairs at the LHC. The calculated cross sections are included in 
Table~\ref{quarkpairs}. For the same reasons as above, the cross sections are about 3-4 
orders of magnitude smaller than for photoproduction. 

The experimental study of ultra-peripheral collisions has so far mainly been concentrated 
to coherent and exclusive interactions, in particular such interactions in which a single 
meson is produced. This can proceed both through a two-photon interaction and through 
coherent photonuclear interactions. The cross sections 
for photonuclear vector meson production are about a factor of 100 higher than the 
corresponding two-photon cross sections\cite{Bertulani:2005ru}.  

The calculation of vector meson production cross 
sections in ultra-peripheral collisions is discussed in \cite{Klein:1999qj,Frankfurt:2002sv}. 
The interest in this process derives mainly from the fact that photoproduction of heavy 
vector mesons is a good probe of the nuclear gluon distribution. 

The total exclusive cross section can be written as an integral over the equivalent photon 
energy 
\begin{equation}
\label{eq:sigma}
\sigma( A + A \rightarrow A + A + V ) = 2 \int \sigma_{\gamma + A \rightarrow V + A} (k) 
\frac{dn_{\gamma}}{dk} dk \; .
\end{equation}
The ``2'' takes into account the fact that both nuclei can act as both target and photon 
emitter. The integral is cut-off at high values of k by the drop in $dn_{\gamma}/dk$, and at low 
values of $k$ by the nuclear form factor or by the threshold energy 
for the process $\gamma + A \rightarrow V + A$. 

\begin{figure} 
\begin{center} 
\includegraphics[width=12cm]{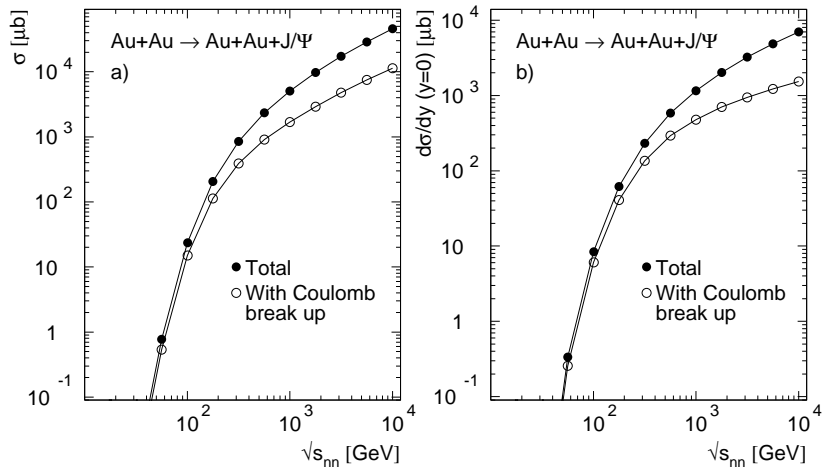} 
\caption{Excitation function for exclusive $J / \Psi$ production in Au+Au interactions. The total 
cross section in shown in a), and the differential cross section ($d \sigma / dy$) at mid-rapidity 
is shown in b). The calculations are based on \cite{Klein:1999qj,Baltz:2002pp}.}
\label{excitation} 
\end{center} 
\end{figure} 

The rapidity, $y$, of the produced vector meson is related to its mass, $M_V$, and the photon energy 
through $k = (M_V/2) \exp(\pm y)$. The $\pm$ corresponds to the two different photon emitter 
and target configurations. The differential form of Eq.~\ref{eq:sigma} is thus
\begin{equation}
\frac{d \sigma(A + A \rightarrow A + A + V)}{dy} = 
k_1 \frac{dn_{\gamma}}{dk_1} \sigma_{\gamma + A \rightarrow V + A} (k_1) +
k_2 \frac{dn_{\gamma}}{dk_2} \sigma_{\gamma + A \rightarrow V + A} (k_2) \; ,
\end{equation}
where $k_1 = (M_V/2) \exp(+y)$ and $k_2 = (M_V/2) \exp(-y)$. At mid-rapidity, $k_1 = k_2$ and 
the contributions from the two terms are equal. Away from $y=0$, there is no obvious way to 
separate the contributions from the two possibilities. 

The excitation function for exclusive $J / \Psi$ production in Pb+Pb interactions is 
shown in Fig.~\ref{excitation}. The calculations are done for exclusive production, and 
for the case where at least one of the nuclei breaks up following Coulomb 
excitation\cite{Baltz:2002pp}. In the latter case, at least one additional photon is exchanged 
that excites one of the nuclei, for example to a Giant Dipole Resonance.

\section{Experimental results on ultra-peripheral collisions}

The limit on the maximum photon energies from the coherence requirement  
discussed in the previous section implies that, before RHIC, 
particle production in ultra-peripheral collisions was in practice 
restricted to two-photon production of $e^+e^-$-pairs. 
This process was indeed studied with heavy-ion beams at the Berkeley 
Bevalac\cite{Belkacem:1997jh}, the BNL AGS\cite{eeAGS}, and the 
CERN SPS\cite{Vane:1992ms}. The results at the AGS and SPS were found to be 
in reasonable agreement with theoretical predictions, while some disagreements were 
reported by the experiment at the Bevalac.  

At RHIC, however, a variety of final states are accessible. 
Coherent production of $\rho^0$--mesons was observed by the STAR 
collaboration shortly after the first collisions at RHIC\cite{Klein:2001vh}. The subsequent 
measurement of the cross section gave a value for the total exclusive production, 
including nuclear break up, $Au+Au \rightarrow Au^{(*)} + Au^{(*)} +\rho^0$, of 
$\sigma =$~460$\pm$220(stat.)$\pm$110(syst.)mb~\cite{Adler:2002sc}, in agreement with 
theoretical estimates\cite{Klein:1999qj,Frankfurt:2002sv}. 
It should be noted that this is nearly 10\% of the 
total $Au+Au$ inelastic cross section! The coherent events were identified from their 
$p_T$ spectrum, which was peaked at very low values $\sim \sqrt{2}/R_{Au} \approx 40$~MeV, in 
sharp contrast to the typical $<p_T>$ in hadronic events ($\approx 350$~MeV). 

The STAR collaboration has also studied production of $e^+e^-$--pairs in two-photon 
interactions\cite{Adams:2004rz}. The results were found to be in good agreement 
with lowest order perturbation theory. The pair--$p_T$ distribution deviated from the 
Weizs\"{a}cker-Williams virtual photon approach, showing that the virtual photon mass 
was important in that kinematic regime. 

The luminosity at RHIC has increased significantly since the first run in the year 2000 
and is now higher than the original design value. 
The increased luminosity has meant that more rare 
processes can been studied. The PHENIX collaboration has thus tried to study exclusive 
$J / \Psi$ production, and the first preliminary results have been 
presented\cite{d'Enterria:2006ep}. The $J / \Psi$s were measured through their decay to 
$e^+e^-$; the electrons and positrons were identified by the PHENIX mid-rapidity Ring Imaging 
Cherenkov Counters and Electromagnetic Calorimeter.
Somewhat surprisingly, the background from two-photon 
production of $e^+e^-$--pairs constituted a significant, but tractable, background. 

The PHENIX result on $J / \Psi$ and high-mass $e^+e^-$--pair production, and the STAR 
results on $e^+e^-$--pair production, were obtained in coincidence with Coulomb 
break up of one or both nuclei by detecting the decay neutrons in Zero-Degree Calorimeters. 
This requirement reduced the trigger background and simplified the normalization. 
STAR has also studied $\rho^0$--meson production with Coulomb break up.

\section{Ultra-peripheral collisions at the LHC}

As was shown in Fig.~1, experiments at the LHC will have the possibility to bring the study 
of photon-proton and photon-nucleus interactions into a new energy domain. Although no LHC 
experiment was designed to investigate such interactions, there are plans for 
studying ultra-peripheral collisions in some of them. 

Two general issues that must be solved are triggering and the separation of a signal 
from background. The 
``background'' can here be both a real background, from e.g. cosmic rays and beam-gas 
interactions, and a ``background'' from hadronic interactions. The triggering will be a 
challenge since some experiments primarily trigger on particle production outside the 
central rapidity region (ALICE), and some require trigger particles with very high  
$p_T$ (ATLAS, and to some extent CMS). 

The ALICE collaboration has included UPCs as a physics topics 
of interest in its Physics Performance Report\cite{ALICE2}. The main focus so far 
has been on coherent production of heavy vector mesons. The $J / \Psi$ and $\Upsilon$ mesons 
can be reconstructed through their decay into di--lepton pairs in the central detectors 
(pseudo-rapidity range $|\eta| < 1$) and in the muon spectrometer ($2.2 \leq \eta \leq 4.0$). 
Electrons can be identified in the central barrel using the Transition Radiation Detector (TRD), 
and muons can be identified in the muon spectrometer. The event reconstruction can 
thus be made using the same detectors and detection techniques as in central collisions. 

The trigger will require separate attention. The principal trigger detector for normal,  
min. bias Pb+Pb events will be the V0 detectors, segmented scintillator counters located 
at pseudorapidities $2.8 \leq \eta \leq 5.1$ and $1.7 \leq  -\eta \leq 3.8$. These will 
not suffice for ultra-peripheral collisions. A few different trigger approaches 
have been investigated\cite{ALICE2}. One is based on using a low-multiplicity signal from the 
Si-Pixel (available at the lowest trigger level) as a pre-trigger for the TRD, which can then 
provide an electron/positron trigger. Alternatively, 
the Time-of-Flight detector can provide a multiplicity trigger at the lowest
trigger level. It might also be possible to use the Zero-Degree Calorimeters 
for triggering on ultra-peripheral events with Coulomb break up. 

There is also an interest in ALICE to study incoherent photon-parton interactions, for example 
production of heavy quark-pairs and photon induced jets. 

\begin{figure}
\begin{center}
\includegraphics[width=7cm]{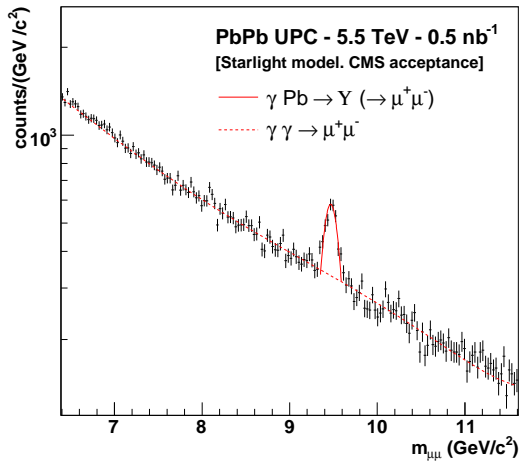}
\caption{Expected $\mu^+ \mu^-$ invariant mass spectrum from coherent 
$\Upsilon \rightarrow \mu^+ \mu^-$ and $\gamma + \gamma \rightarrow \mu^+ \mu^-$ interactions 
in the CMS acceptance for a $10^6$~s heavy-ion run at the LHC. 
From~\cite{d'Enterria:2006ve}.}
\label{dde}
\end{center}
\end{figure}

The possibility to study UPCs in the CMS experiment has been investigated\cite{d'Enterria:2006ve}. 
Also in CMS, most attention so far has been  
directed towards heavy vector mesons, in particular coherent $\Upsilon$ production 
reconstructed through the decay channel $\Upsilon \rightarrow \mu^+ \mu^-$. 
The $\Upsilon$ is chosen rather than the $J / \Psi$ since the muons from the decay have 
to be energetic enough to reach the muon chambers. The simulated $\mu^+ \mu^-$ invariant
mass spectrum from a 0.5 nb$^{-1}$ run (the expected integrated luminosity for a standard 
LHC month of running with heavy ions) is shown in Fig.~\ref{dde}. A $\Upsilon$ peak can 
be clearly identified above the continuum. The statistics will be sufficient for a meaningful 
measurement. 

The $J / \Psi$ and also the $\Upsilon$ can be reconstructed through the decay into 
$e^+ e^-$-pairs by using the CMS Electromagnetic Calorimeters. 

The proposal to equip ATLAS with Zero-Degree Calorimeters\cite{SWhite} could make it possible to 
study ultra-peripheral collisions in ATLAS.

\section{Conclusion}

Ultra-peripheral collisions can enrich the physics potential of experiments at 
hadron colliders. Although the cross sections for photon-induced processes are normally 
small in comparison with those for the corresponding processes in hadronic interactions, 
the feasibility of extracting a signal for at least a few interesting reaction channels 
has been shown by experiments at RHIC. The equivalent 
photon energies at the LHC will be higher than at any existing accelerator; this is true 
both in proton-proton and nucleus-nucleus collisions. 

\section*{Acknowledgements} 
I have, over the years, benefited from discussions on the topics of this talk 
with several people. A few I would like to mention are 
A. Baltz, G. Baur, C. Bertulani, D. d'Enterria, S. Klein, M. Strikman, D. Silvermyr, 
R. Vogt, and S. White.

\end{document}